\documentclass[pdflatex,12pt]{article}
\usepackage{graphicx}

\def\pbnr{}
\def\speaker{Patrick Spradlin}
\def\onbehalfof{the \LHCb collaboration}
\def\title{The \LHCb prompt charm triggers}
\def\affiliation{School of Physics and Astronomy\\
University of Glasgow, Glasgow, UK}
\def\support{}

\usepackage{xspace}
\usepackage{lineno}
\usepackage{amsmath}
\usepackage{ifthen}
\newboolean{uprightparticles}
\setboolean{uprightparticles}{false}
\newboolean{articletitles}
\setboolean{articletitles}{true} 
\usepackage{amssymb}
\usepackage{amsfonts}
\usepackage{upgreek}

\usepackage{caption}
\usepackage{subfig}

\usepackage{color}
\usepackage{wrapfig}
\usepackage{mciteplus}

\usepackage{tabularx}
\usepackage{dcolumn}
\newcolumntype{d}[1]{D{.}{.}{#1}}
\newcolumntype{C}{>{\centering}X}
\usepackage{calc}

\usepackage{hyperref}    

\newcommand\pubnumber{\pbnr}
\newcommand\pubdate{\today}
\def\Title#1{\begin{center} {\Large #1 } \end{center}}
\def\Author#1{\begin{center}{ \sc #1} \end{center}}

\newcommand{\OnBehalf}[1]{\sbox0{#1}\ifdim\wd0=0pt
        {}
	\else
	{\\on behalf of #1}
	\fi}
\newcommand{\SupportedBy}[1]{\sbox0{#1}\ifdim\wd0=0pt
        {}
	\else
	{\footnote{#1}}
	\fi}
\def\Address#1{\begin{center}{ \it #1} \end{center}}

\newcommand\pubblock{\includegraphics[width=5cm]{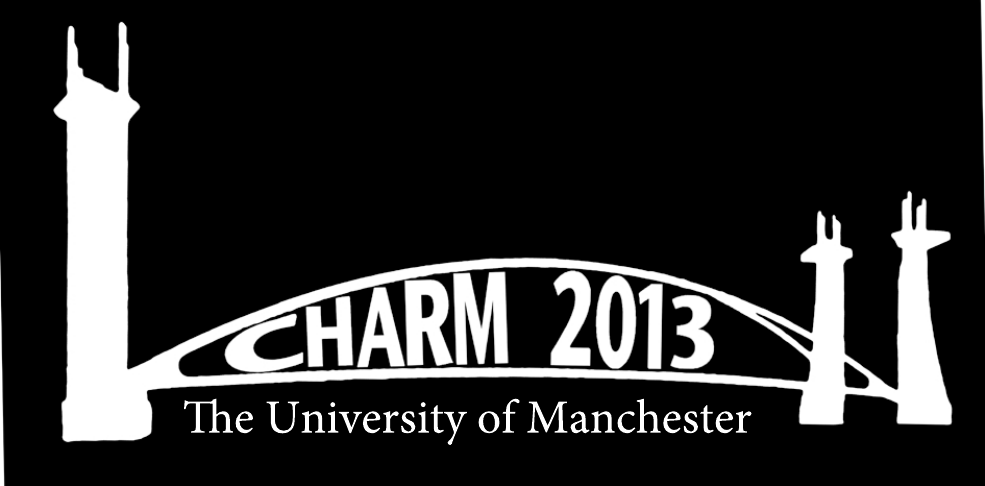}\hfill{\begin{tabular}{l} \pubnumber\\
         \pubdate  \end{tabular}}}
\newenvironment{Abstract}{\begin{quotation}  }{\end{quotation}}
\newenvironment{Presented}{\begin{quotation} \begin{center} 
             PRESENTED AT\end{center}\bigskip 
      \begin{center}\begin{large}}{\end{large}\end{center} \end{quotation}}

\def\venue{The 6$^{th}$ International Workshop on Charm Physics\\
(CHARM 2013)\\
Manchester, UK,  31 August -- 4 September, 2013}

\def\beq{\begin{equation}}
\def\eeq#1{\label{#1}\end{equation}}
\def\eeqn{\end{equation}}

\def\beqa{\begin{eqnarray}}
\def\eeqa#1{\label{#1}\end{eqnarray}}
\def\eeqan{\end{eqnarray}}

\let\bar=\overbar

\def\eg{{\it e.g.}}

\def\D{{\cal D}}

\def\Dslash{\not{\hbox{\kern-4pt $D$}}}
\def\dslash{\not{\hbox{\kern-2pt $\del$}}}

\def\msb{{\bar{\ssstyle M \kern -1pt S}}}

\def\lhcb {\mbox{LHCb}\xspace}
\def\ux85 {\mbox{UX85}\xspace}

\def\lhc    {\mbox{LHC}\xspace}

\def\lone   {L0\xspace}
\def\hlt    {HLT\xspace}
\def\hltone {HLT1\xspace}
\def\hlttwo {HLT2\xspace}

\ifthenelse{\boolean{uprightparticles}}%
{

 \def\Ppi         {\ensuremath{\uppi}\xspace}

 \def\Ppsi        {\ensuremath{\uppsi}\xspace}

 \def\PDelta      {\ensuremath{\Delta}\xspace}                 
 \def\PXi      {\ensuremath{\Xi}\xspace}                 
 \def\PLambda      {\ensuremath{\Lambda}\xspace}                 
 \def\PSigma      {\ensuremath{\Sigma}\xspace}                 
 \def\POmega      {\ensuremath{\Omega}\xspace}                 
 \def\PUpsilon      {\ensuremath{\Upsilon}\xspace}                 
 
 \def\PB      {\ensuremath{\mathrm{B}}\xspace}                 
                  
 \def\PD      {\ensuremath{\mathrm{D}}\xspace}

 \def\PJ      {\ensuremath{\mathrm{J}}\xspace}                 
 \def\PK      {\ensuremath{\mathrm{K}}\xspace}

 \def\Pb      {\ensuremath{\mathrm{b}}\xspace}                 
 \def\Pc      {\ensuremath{\mathrm{c}}\xspace}

 \def\Ph      {\ensuremath{\mathrm{h}}\xspace}                 
 \def\Pi      {\ensuremath{\mathrm{i}}\xspace}

 \def\Pp      {\ensuremath{\mathrm{p}}\xspace}

}
{

 \def\Ppi         {\ensuremath{\pi}\xspace}

 \def\Ppsi        {\ensuremath{\psi}\xspace}                 
                  
 \mathchardef\PDelta="7101
 \mathchardef\PXi="7104
 \mathchardef\PLambda="7103
 \mathchardef\PSigma="7106
 \mathchardef\POmega="710A
 \mathchardef\PUpsilon="7107
                  
 \def\PB      {\ensuremath{B}\xspace}                 
                  
 \def\PD      {\ensuremath{D}\xspace}

 \def\PJ      {\ensuremath{J}\xspace}                 
 \def\PK      {\ensuremath{K}\xspace}

 \def\Pb      {\ensuremath{b}\xspace}                 
 \def\Pc      {\ensuremath{c}\xspace}

 \def\Ph      {\ensuremath{h}\xspace}                 
 \def\Pi      {\ensuremath{i}\xspace}

 \def\Pp      {\ensuremath{p}\xspace}

}

\def\cquark    {\ensuremath{\Pc}\xspace}
\def\cquarkbar {\ensuremath{\overline \cquark}\xspace}
\def\ccbar     {\ensuremath{\cquark\cquarkbar}\xspace}
\def\bquark    {\ensuremath{\Pb}\xspace}
\def\bquarkbar {\ensuremath{\overline \bquark}\xspace}
\def\bbbar     {\ensuremath{\bquark\bquarkbar}\xspace}

\def\pion  {\ensuremath{\Ppi}\xspace}

\def\pip   {\ensuremath{\pion^+}\xspace}
\def\pim   {\ensuremath{\pion^-}\xspace}

\def\kaon  {\ensuremath{\PK}\xspace}
  \def\Kbar  {\kern 0.2em\overline{\kern -0.2em \PK}{}\xspace}

\def\Kz    {\ensuremath{\kaon^0}\xspace}
\def\Kzb   {\ensuremath{\Kbar^0}\xspace}
\def\KzKzb {\ensuremath{\Kz \kern -0.16em \Kzb}\xspace}
\def\Kp    {\ensuremath{\kaon^+}\xspace}
\def\Km    {\ensuremath{\kaon^-}\xspace}

\def\KpKm  {\ensuremath{\Kp \kern -0.16em \Km}\xspace}

  \def\Dbar    {\kern 0.2em\overline{\kern -0.2em \PD}{}\xspace}
\def\D       {\ensuremath{\PD}\xspace}

\def\Dz      {\ensuremath{\D^0}\xspace}
\def\Dzb     {\ensuremath{\Dbar^0}\xspace}
\def\DzDzb   {\ensuremath{\Dz {\kern -0.16em \Dzb}}\xspace}
\def\Dp      {\ensuremath{\D^+}\xspace}
\def\Dm      {\ensuremath{\D^-}\xspace}

\def\DpDm    {\ensuremath{\Dp {\kern -0.16em \Dm}}\xspace}

\def\Dstarp  {\ensuremath{\D^{*+}}\xspace}

\def\B       {\ensuremath{\PB}\xspace}
  \def\Bbar    {\kern 0.18em\overline{\kern -0.18em \PB}{}\xspace}

\def\Bz      {\ensuremath{\B^0}\xspace}

\def\Bu      {\ensuremath{\B^+}\xspace}

\def\Bp      {\ensuremath{\Bu}\xspace}

\def\jpsi     {\ensuremath{{\PJ\mskip -3mu/\mskip -2mu\Ppsi\mskip 2mu}}\xspace}

  \def\Y#1S{\ensuremath{\PUpsilon{(#1S)}}\xspace}

\def\proton      {\ensuremath{\Pp}\xspace}

\def\Lbar {\ensuremath{\kern 0.1em\overline{\kern -0.1em\PLambda}}\xspace}
\def\Lambdares {\ensuremath{\PLambda}\xspace}

\newcommand{\decay}[2]{\mbox{\ensuremath{#1\!\to #2}}\xspace}   

\def\to                 {\ensuremath{\rightarrow}\xspace}

\def\CP                {\ensuremath{C\!P}\xspace}

\def\AT#1     {\ensuremath{A_{\mathrm{T}}^{#1}}\xspace}           

\def\C#1      {\ensuremath{\mathcal{C}_{#1}}\xspace}                       
\def\Cp#1     {\ensuremath{\mathcal{C}_{#1}^{'}}\xspace}                    
\def\Ceff#1   {\ensuremath{\mathcal{C}_{#1}^{\mathrm{(eff)}}}\xspace}        
\def\Cpeff#1  {\ensuremath{\mathcal{C}_{#1}^{'\mathrm{(eff)}}}\xspace}       
\def\Ope#1    {\ensuremath{\mathcal{O}_{#1}}\xspace}                       
\def\Opep#1   {\ensuremath{\mathcal{O}_{#1}^{'}}\xspace}                    

\newcommand{\tev}{\ensuremath{\mathrm{\,Te\kern -0.1em V}}\xspace}
\newcommand{\gev}{\ensuremath{\mathrm{\,Ge\kern -0.1em V}}\xspace}
\newcommand{\mev}{\ensuremath{\mathrm{\,Me\kern -0.1em V}}\xspace}
\newcommand{\kev}{\ensuremath{\mathrm{\,ke\kern -0.1em V}}\xspace}
\newcommand{\ev}{\ensuremath{\mathrm{\,e\kern -0.1em V}}\xspace}
\newcommand{\gevc}{\ensuremath{{\mathrm{\,Ge\kern -0.1em V\!/}c}}\xspace}
\newcommand{\mevc}{\ensuremath{{\mathrm{\,Me\kern -0.1em V\!/}c}}\xspace}
\newcommand{\gevcc}{\ensuremath{{\mathrm{\,Ge\kern -0.1em V\!/}c^2}}\xspace}
\newcommand{\gevgevcccc}{\ensuremath{{\mathrm{\,Ge\kern -0.1em V^2\!/}c^4}}\xspace}
\newcommand{\mevcc}{\ensuremath{{\mathrm{\,Me\kern -0.1em V\!/}c^2}}\xspace}

\def\mm   {\ensuremath{\rm \,mm}\xspace}

\def\mum  {\ensuremath{\,\upmu\rm m}\xspace}

\def\mub{\ensuremath{\rm \,\upmu b}\xspace}

\def\ns   {\ensuremath{{\rm \,ns}}\xspace}

\def\mhz  {\ensuremath{{\rm \,MHz}}\xspace}
\def\khz  {\ensuremath{{\rm \,kHz}}\xspace}

\def\gsim{{~\raise.15em\hbox{$>$}\kern-.85em
          \lower.35em\hbox{$\sim$}~}\xspace}
\def\lsim{{~\raise.15em\hbox{$<$}\kern-.85em
          \lower.35em\hbox{$\sim$}~}\xspace}

\def\pt         {\mbox{$p_{\rm T}$}\xspace}
\def\et         {\mbox{$E_{\rm T}$}\xspace}

\def\tell1  {TELL1\xspace}
\def\ukl1   {UKL1\xspace}

\newcommand{\LHCb}{{\upshape{LHCb}}\xspace}

\def\Dpors      {\ensuremath{\D^+_{(s)}}\xspace}

\def\Lcp     {\ensuremath{\Lambdares_\cquark^+}\xspace}

\newcommand{\TeV}{\tev}
\newcommand{\GeV}{\gev}

\newcommand{\keV}{\keV}

\newcommand{\GeVc}{\gevc}
\newcommand{\MeVc}{\mevc}

\def\MHz  {\mhz}
\def\kHz  {\khz}

\def\pT         {\pt}

\newcommand{\lonehad}{\texttt{L0Hadron}\xspace}
\newcommand{\hltonetrack}{\texttt{Hlt1TrackAllL0}\xspace}

\newcommand{\hlttwocharmtwobody}{\texttt{Hlt2CharmHadD02HH\_D02KPi}\xspace}
\newcommand{\hlttwocharmthreebody}{\texttt{Hlt2CharmHadD2HHH}\xspace}
\newcommand{\hlttwocharmfourbody}{\texttt{Hlt2CharmHadD02HHHHDst\_K3pi}\xspace}
\newcommand{\hlttwocharmincldstar}{\texttt{Hlt2CharmHadD02HHXDst\_hhX}\xspace}

\newcommand{\DzToKmpip}{\decay{\Dz}{\Km \pip}}
\newcommand{\DzToKmpippippim}{\decay{\Dz}{\Km \pip\pip\pim}}

\newcommand{\DpToKmpippip}{\decay{\Dp}{\Km \pip\pip}}

\newcommand{\DstarpTopipDz}{\decay{\Dstarp}{\pip\Dz}}

\newcommand{\DstarpTopipDzToKmpippippim}{\decay{\Dstarp}{\pip \Dz(\Km \pip \pip \pim)}}

\newcommand{\TOS}{\ensuremath{\mathrm{TOS}}\xspace}
\newcommand{\TIS}{\ensuremath{\mathrm{TIS}}\xspace}

\newcommand{\efftos}{\ensuremath{\epsilon^{\TOS}}\xspace}
\newcommand{\ntis}{\ensuremath{N^{\TIS}}\xspace}
\newcommand{\ntistos}{\ensuremath{N^{\TOS \wedge \TIS}}\xspace}

\newcommand{\Tabref}[1]{Table~\ref{#1}\xspace}

\newcommand{\Figref}[1]{Figure~\ref{#1}\xspace}

\newcommand{\Secref}[1]{Section~\ref{#1}\xspace}
\newcommand{\secref}[1]{Sec.~\ref{#1}\xspace}

\newcommand{\refref}[1]{Ref.~\cite{#1}\xspace}

\newcommand{\Secsref}[2]{Sections~\ref{#1} to~\ref{#2}\xspace}

\newcommand{\xsecstxtnoun}{cross-sections\xspace}

\usepackage[defaultlines=3,all]{nowidow}

\begin{document}
\begin{titlepage}
\pubblock

\vfill
\Title{\title}
\vfill
\Author{\speaker\SupportedBy{\support}\OnBehalf{\onbehalfof}}
\Address{\affiliation}
\vfill
\begin{Abstract}
  The \LHCb experiment has fully reconstructed close to $10^{9}$ charm hadron
  decays---by far the world's largest sample.
  During the 2011-2012 running periods, the effective $\proton\proton$ beam
  crossing rate was 11-15\MHz while the rate at which events were written to
  permanent storage was 3-5\kHz.
  Prompt charm candidates (produced at the primary interaction vertex) were
  selected using a combination of exclusive and inclusive high level
  (software) triggers in conjunction with low level hardware triggers.
  The efficiencies, background rates, and possible biases of the triggers
  as they were implemented will be discussed, along with plans for the running
  at 13\TeV in 2015 and subsequently in the upgrade era.
\end{Abstract}
\vfill
\begin{Presented}
\venue
\end{Presented}
\vfill
\end{titlepage}
\def\thefootnote{\fnsymbol{footnote}}
\setcounter{footnote}{0}
%


\section{Introduction}
\label{sec:intro}

  The \LHCb experiment has rapidly become one of the foremost high-precision
  flavor physics experiments, collecting the world's largest samples of
  several decay modes of \cquark and
  \bquark-hadrons (\eg \cite{LHCb-PAPER-2013-053,LHCb-PAPER-2013-037}).
  This success would have been impossible without \LHCb's flexible and
  efficient trigger system.

  The task of rapidly selecting which events will be stored permanently for
  subsequent analysis and which will be discarded
  forever---triggering---presents a formidable challenge in the high-energy
  hadronic collision environment of the Large Hadron Collider (\lhc).
  In 2012 the \LHCb detector witnessed \mbox{$\proton\proton$} collisions with a
  center-of-mass energy of \mbox{$\sqrt{s} = 8\TeV$} at a mean
  instantaneous luminosity of approximately
  \mbox{$4 \times 10^{32}\,\mathrm{cm}^{-2}\mathrm{s}^{-1}$}.
  Given that the heavy flavor hadron production \xsecstxtnoun 
  into the \LHCb acceptance were measured to be
  $\sigma_{\bbbar,\mathrm{acc}} = 75.3 \pm 14.1 \mub$~\cite{LHCb-PAPER-2010-002} and
  $\sigma_{\ccbar,\mathrm{acc}} = 1419 \pm 134\mub$~\cite{LHCb-PAPER-2012-041}
  for \mbox{$\proton\proton$} collisions at \mbox{$\sqrt{s} = 7\TeV$}, the
  rate of heavy flavor production into the \LHCb acceptance exceeded 30\kHz
  for \bquark-hadrons and 600\kHz for \cquark-hadrons.
  Because events are written to permanent storage at just \mbox{3-5\kHz}, the
  trigger must be highly selective even among events with a real heavy-flavor
  hadron.

  This article discusses the structure and performance of the trigger
  components for selecting events that contain open charm hadrons---the first
  and fundamental building block for most precision charm measurements at
  \LHCb.
  We also sketch prospective improvements to the trigger that will
  extend our physics reach when the \lhc returns to operation after its first
  long shutdown period (LS1) and in the era of the upgraded \LHCb detector.

\section{Detector}
\label{sec:det}

  The \LHCb detector~\cite{Alves:2008zz} is a single-arm forward
  spectrometer covering the \mbox{pseudorapidity} range \mbox{$2 < \eta <5$},
  designed for the study of particles containing \bquark or \cquark
  quarks.
  Charm hadron triggering uses information from each of the detector
  subsystems.
  The detector includes a high-precision tracking system
  consisting of a silicon-strip vertex detector surrounding the
  \mbox{$\proton\proton$} interaction region, a large-area silicon-strip
  detector located upstream of a dipole magnet with a bending power of about
  $4{\rm\,Tm}$, and three stations of silicon-strip detectors and straw
  drift tubes placed downstream.
  The combined tracking system provides a momentum measurement with
  relative uncertainty that varies from 0.4\% at 5\gevc to 0.6\% at 100\gevc,
  and impact parameter resolution of 20\mum for tracks with large transverse
  momentum.
  Different types of charged hadrons are distinguished by information
  from two ring-imaging Cherenkov detectors~\cite{LHCb-DP-2012-003}.
  Photon, electron, and hadron candidates are identified by a calorimeter
  system consisting of scintillating-pad and preshower detectors, an
  electromagnetic calorimeter and a hadronic calorimeter.
  Muons are identified by a system composed of alternating layers of iron and
  multiwire proportional chambers~\cite{LHCb-DP-2012-002}.

\section{Trigger overview}
\label{sec:over}

  \begin{wrapfigure}{R}{0.4\textwidth}
    \centering
     \includegraphics[type=pdf,ext=.pdf,read=.pdf,width=0.39\textwidth]{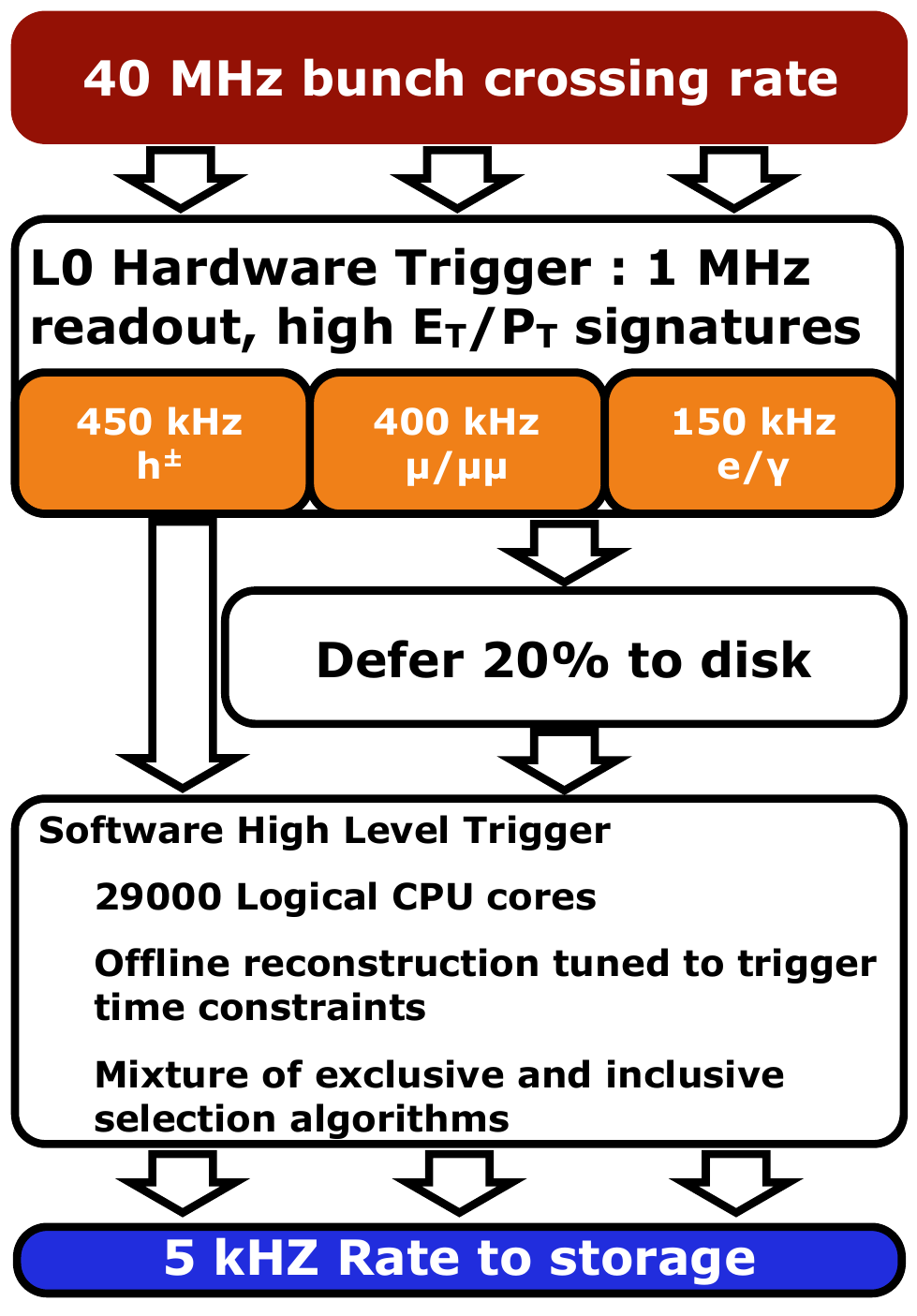}
  \caption{\small A diagrammatic overview of the trigger structure.
   \label{fig:over:over}}
  \end{wrapfigure}

  Although the global structure of the \LHCb trigger system---a hardware
  trigger system followed by a full detector readout and one or more layers of
  software triggers---has remained unchanged since its initial
  design~\cite{LHCbTDR:2003tg}, the implementation continues to evolve.
  The trigger system as it performed in 2011 is described in detail in
  \refref{LHCb-DP-2012-004}, but the interval between 2011 and 2012 saw the
  introduction of a major new feature, \hlt deferral (\secref{sec:defer}).
  The steady evolution of the trigger has led to and has been encouraged by
  an expansion of \LHCb's physics program.
  Relative to the initial design, the 2012 \LHCb trigger processed twice
  the instantaneous luminosity of events
  (\mbox{$4 \times 10^{32}\,\mathrm{cm}^{-2}\mathrm{s}^{-1}$}
  vs.\ \mbox{$2 \times 10^{32}\,\mathrm{cm}^{-2}\mathrm{s}^{-1}$})
  with a much greater complexity (a mean of $1.6$ visible
  \mbox{$\proton\proton$} interactions per visible bunch crossing
  vs.\ $0.4$) and recorded events to permanent storage at over twice the
  rate (5\kHz vs.\ 2\kHz).
  As a consequence, \LHCb is making an impact in areas far outside its
  initial core physics program~\cite{Adeva:2009ny,LHCb-PAPER-2012-031},
  particularly in the realm of charm physics.
  Though charm physics measurements were previously absent from \LHCb's primary
  goals, approximately $40\%$ of the trigger output is now dedicated to them.

  \Figref{fig:over:over} outlines the structure of the trigger system for
  2012 data collection.
  The chain begins with a bunch crossing in which a bunch of protons from
  each of the counter-rotating beams of the \lhc meet at the \LHCb interaction
  point.
  The separation between successive potential sites for bunches of protons
  in the beams of the \lhc  is $25\ns$, thus bunch crossings may occur at
  a maximum rate of 40\MHz~\cite{Evans:2008zzb}.
  In much of 2012 the actual bunch crossing rate at the \LHCb interaction
  point was 11-15\MHz.

  The first layer of triggering happens in bespoke hardware.
  Since the maximum rate at which the full detector response can be
  digitized and read out is 1\MHz, the purpose of this level-0 trigger
  system (\lone) is to select just 1\MHz of potentially interesting events
  from the 11-15\MHz of bunch crossings.
  \lone analyzes the response of selected subdetectors
  to evaluate measures of event complexity and
  to identify signatures of particles with large momentum components
  transverse to the \mbox{$\proton\proton$} collision axis (\pT).
  It contains a number of independent parallel configurable channels that
  are tuned to balance the requirements of the physics program and the
  readout constraint.
  If any one of the channels returns a positive decision, the full detector
  response is digitized, read out, and recorded to the temporary storage
  of the Event Filter Farm (EFF), a large farm of multiprocessor computers,
  until the trigger processing is complete and a final decision made on the
  fate of the event.

  Most of the events accepted by \lone and transferred to the EFF are
  processed immediately by the subsequent and final triggering layer, the
  High Level Trigger (\hlt).
  For 20\% of the events the \hlt processing is deferred until the interfill
  period (see \secref{sec:defer}).
  \hlt is implemented in software that runs on the EFF.
  Due to limitations of computing resources available for permanent storage
  and data analysis, the rate at which events are accepted for permanent
  storage is restricted to 5\kHz.

  Internally, \hlt is segmented into two sequential stages of processing,
  \hltone and \hlttwo.
  Each stage is composed of several independent parallel channels (lines)
  that are sequences of event reconstruction algorithms and selection criteria.
  Each line executes its sequence of elements either until the decision of
  the line is known to be negative, \eg, by the failure of a reconstruction
  element or selection criterion, or until the sequence is complete and the
  event accepted by the line.
  The lines of \hlttwo are executed only for events that are
  accepted by at least one of the lines of \hltone.
  Events accepted by at least one \hlttwo line are preserved in permanent
  storage.

  The lines of \hltone are simple selections based on the properties of one
  or two reconstructed tracks.
  The lines of \hlttwo can be quite sophisticated, incorporating complicated
  reconstruction elements and multivariate discriminants, and are generally
  tailored to the requirements of a group of physics analyses.
  The lines of \hlttwo are generally better suited to the needs of \LHCb
  measurements than those of \hltone.
  However, they also require substantial computing resources.
  The EFF has the computing power to execute the lines of \hlttwo on only
  a fraction of the \lone-accepted events.
  Thus the two-stage structure of \hlt is a compromise, with \hltone rapidly
  selecting a subset of the \lone-accepted events to be further analyzed
  by \hlttwo.

\section{\hlt deferral}
\label{sec:defer}

  The trigger system in 2012 featured a new facility for deferring
  \hlt processing for a fraction of the events accepted by \lone.
  This represents a significant improvement in the efficiency with which
  the EFF is used.
  Prior to the implementation of \hlt deferral, all events were processed
  immediately after they were transferred to the EFF.
  In normal operation, the beams of the \lhc are dumped when their intensity
  decays below some threshold.
  New beams with renewed intensity are then injected and accelerated to the
  target energy before collisions resume.
  This interfill period in which no recordable collisions occur can take
  a few hours during which the EFF would remain largely idle.
  With the \hlt deferral system, most events are processed immediately, as
  before, but a configurable fraction of the incoming events are cached
  in EFF storage instead of processed.
  During the interfill period, \hlt processes these cached events.
  The net result is a more efficient use of the EFF that effectively increased
  the available computing power by approximately 20\% in 2012.

\section{Performance measures}
\label{sec:tistos}

  We measure the performance of trigger lines in data with the method
  described in \refref{LHCb-DP-2012-004}.
  The data sets for the measurements are collections of `offline' candidate
  decays that have been reconstructed by \LHCb's analysis software from the
  collected events.
  We require that at least one channel at each level of the trigger 
  accepted each event independently of the offline candidate in order to
  mitigate biases due to the \textit{de facto} triggering of the events.

  In order to measure the efficiency with which these offline candidate
  decays satisfy the criteria of a trigger line under investigation, we
  must compare the underlying information from the detector that was used
  in reconstructing the offline candidate to that used in the decision
  of the trigger line.
  This is done by a direct comparison of the set of detector
  elements---the strips, straws, cells, and pads of the
  sub-detectors---that contributed to each.
  As most \hltone and \hlttwo lines are based on sets of reconstructed tracks,
  this is effectively a comparison of the set of tracks constituting the
  offline candidate decay and the set of tracks used by the line.
  We classify an offline candidate as Triggered On Signal (\TOS) for a given
  trigger line if the set of detector elements that was used in its
  reconstruction is sufficient to satisfy the selection criteria of that line.
  An offline candidate is classified as Triggered Independently of Signal
  (\TIS) for a given trigger line if the set of detector elements that was
  used in its reconstruction is disjoint with at least one of the combinations
  of elements that led to a positive decision by that trigger line, that is, if
  the rest of the event excluding the offline signal candidate was sufficient
  to satisfy the criteria of that line.
  These are not mutually exclusive classifications.
  A given offline candidate decay can be both \TOS and \TIS with respect to a
  given trigger line as there may be multiple sets of detector elements 
  whose response led to a positive decision for the line.

  The offline candidate decays of the data sets for trigger performance are
  \TIS with respect to at least one physics line at each level of
  the trigger.
  The candidates of these data sets are largely unbiased by the trigger line
  under investigation.
  A subset of these candidates will also be \TOS with respect to the target
  line.
  After determining the number of signal decays in the set of \TIS candidates
  (\ntis) and of its \TOS subset (\ntistos), we define our measure of the
  performance of a line as its \TOS efficiency,
  \mbox{$\efftos = \ntistos / \ntis$}.

  The \TOS efficiency defined in this way should be considered a relative
  measure of performance rather than an absolute efficiency.
  It is sensitive to the criteria with which the set of offline candidate
  decays were selected.
  Further, the \TIS classification includes some bias due to the pairwise
  production mechanisms of heavy hadrons.
  Despite these limitations, \efftos is an excellent measure of the relative
  performance of a trigger line.

  In \Secsref{sec:lzero}{sec:hlt2} we will show \efftos for offline
  reconstructed decays of three charmed hadrons to final states involving
  kaons and pions:
  \DzToKmpip, \DpToKmpippip, and \DstarpTopipDzToKmpippippim.
  The corresponding charge-conjugate decays are implied here and throughout
  the remainder of this article.
  These modes were selected due to their large abundance and in order to
  show the dependence of trigger efficiencies on the multiplicity of
  the final state.
  Rare open charm hadron decays to final states with two muons are expected
  to have a significantly better performance, comparable to that of
  \jpsi decays (see \refref{LHCb-DP-2012-004}).
  However, their \efftos performance cannot be evaluated until sufficiently
  large samples are available.
  All of the plots and performance estimates in the following sections are
  based on data collected by \LHCb in 2012 in \mbox{$\proton\proton$}
  collisions at $\sqrt{s} = 8\TeV$.

\section{\lone performance}
\label{sec:lzero}

  \begin{wrapfigure}{R}{0.5\textwidth}
    \centering
     \includegraphics[type=pdf,ext=.pdf,read=.pdf,width=0.49\textwidth]{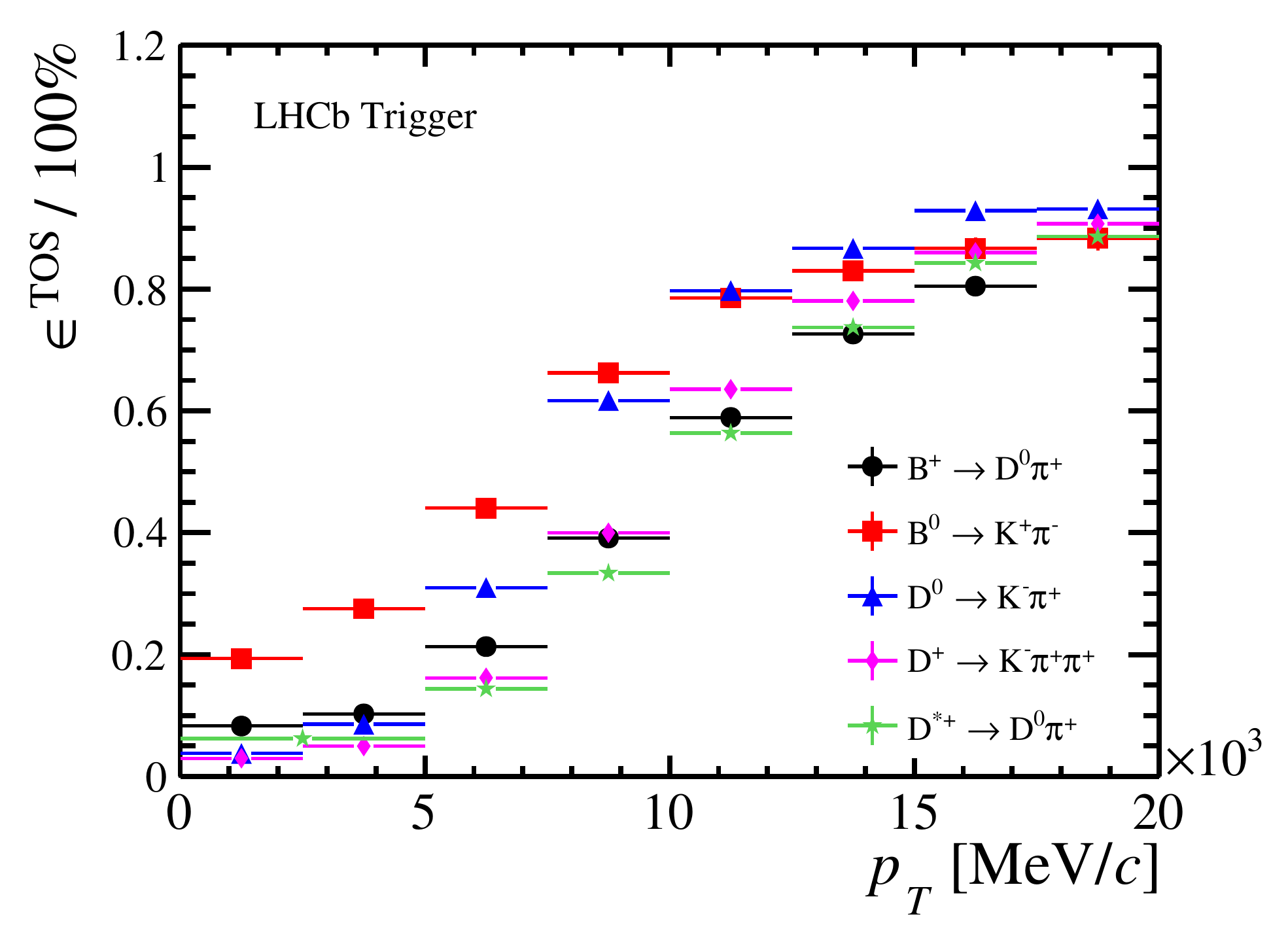}
  \caption{\small The efficiencies of \lonehad for various reconstructed decay
    modes as functions of \pT of the signal \PB and \PD candidate
    based on $\sqrt{s} = 8\TeV$ data collected in 2012.
   \label{fig:lzero:etos}}
  \end{wrapfigure}

  \begin{table}[tp]
  \caption{\small Mean \efftos efficiencies of \lonehad for selected charm
    hadron decays.
   \label{tab:lzero:etos}}
    \centering
      \begin{tabular}{l|d{5}@{\,$\pm$\,}d{5}}
        Decay mode & \multicolumn{2}{c}{Mean \efftos} \\ \hline
        \DzToKmpip                      & 0.26894 & 0.00069 \\
        \DpToKmpippip                   & 0.15766 & 0.00016 \\
        \DstarpTopipDzToKmpippippim     & 0.22045 & 0.00043 \\
      \end{tabular}
  \end{table}

  The \lone hardware trigger system is described more completely in
  Refs.~\cite{Alves:2008zz,LHCb-DP-2012-004}.
  The decisions of the parallel channels are based on comparisons of a
  small number of estimated quantities to specified configurable thresholds.
  The primary physics quantities are the estimated transverse momenta for
  track segments in the muon system and estimated transverse energy
  (\et)\footnote{For a calorimeter cell centered at polar coordinates
  $\vec{x} = (r, \theta, \phi)$ in the \LHCb
  coordinate system in which the origin is at the center of the 
  \mbox{$\proton\proton$} interaction envelope and the $z$-axis is the
  laboratory-frame collision axis, a measured deposited energy of $E$
  corresponds to $\et = E \sin{\theta}$.}
  for clusters in the calorimeter system.
  The overall activity in the scintillating-pad detector enters many \lone
  channels as a proxy measure of event complexity.

  The primary channel of interest for hadronic decays of charmed hadrons
  is the single-cluster hadron line \lonehad.
  It accepts events that have a scintillating-pad detector activity below
  a certain threshold and that contain at least one cluster in the hadron
  calorimeter that has a total transverse energy in all calorimeters of
  \mbox{$\et > 3.5\GeV$}.
  In 2012, approximately 45\% of the events accepted by \lone were accepted
  by \lonehad.
  \Figref{fig:lzero:etos} shows \efftos of \lonehad as a function of signal
  hadron \pT for
  \DzToKmpip, \DpToKmpippip, and \DstarpTopipDzToKmpippippim decays.
  It also shows \efftos of \lonehad for two hadronic \PB decay modes,
  \decay{\Bp}{\pip\Dz(\Km\pip)} and \decay{\Bz}{\Kp\pim}.
  The efficiencies of \lonehad are strongly dependent on the \pT of the
  signal hadron.
  Charm hadrons are predominantly produced in the region of low
  efficiency~\cite{LHCb-PAPER-2012-041}, thus the mean efficiency for the
  set of offline candidate decays is correspondingly low, as shown in
  \Tabref{tab:lzero:etos}.
  One of the important ways in which the redesigned trigger for the upgraded
  \LHCb detector will benefit \LHCb's charm physics program by removing the
  limitations of the \lone system (see \Secref{sec:future:upgrade}).

\section{\hltone performance}
\label{sec:hlt1}

  \hltone, the initial stage of the \hlt software trigger, is composed of
  parallel independent lines---sequences of processing steps that
  include reconstruction elements and selection criteria.
  The decisions of the \lone channels are available to \hltone lines, so
  the trigger history of an event can enter the decision-making process of a
  line.

  Although the lines of \hltone are independent, most lines begin with
  a fast reconstruction of \mbox{$\proton\proton$} primary interaction
  vertexes (PVs) and charged particle tracks that is common to all lines
  that use it.
  The details of this fast reconstruction are fully described in
  \refref{LHCb-DP-2012-004}.
  Most \hltone lines are simple selections based on the properties of one
  or two of these reconstructed tracks.
  The single displaced-track line \hltonetrack, which is the primary \hltone
  line of interest for charmed hadron decays to hadronic final states, is of
  this type.
  \hltonetrack accepts events that were accepted by any \lone channel and
  that have at least one track that satisfies a number of track quality
  criteria (see \refref{LHCb-DP-2012-004}), that is displaced from every
  reconstructed PV in the event (impact parameter with respect to each PV
  $> 0.1\mm$), and that has a relatively large estimated \pT
  ($\pT > 1.7\GeVc$).
  Such tracks are typically produced by the decay products of \cquark and
  \bquark-hadrons and are excellent signatures of long-lived heavy hadrons.

  \begin{figure}[tb]
    \centering
    \subfloat{\label{fig:hlt1:etos:pt}%
      \includegraphics[type=pdf,ext=.pdf,read=.pdf,width=0.49\textwidth]{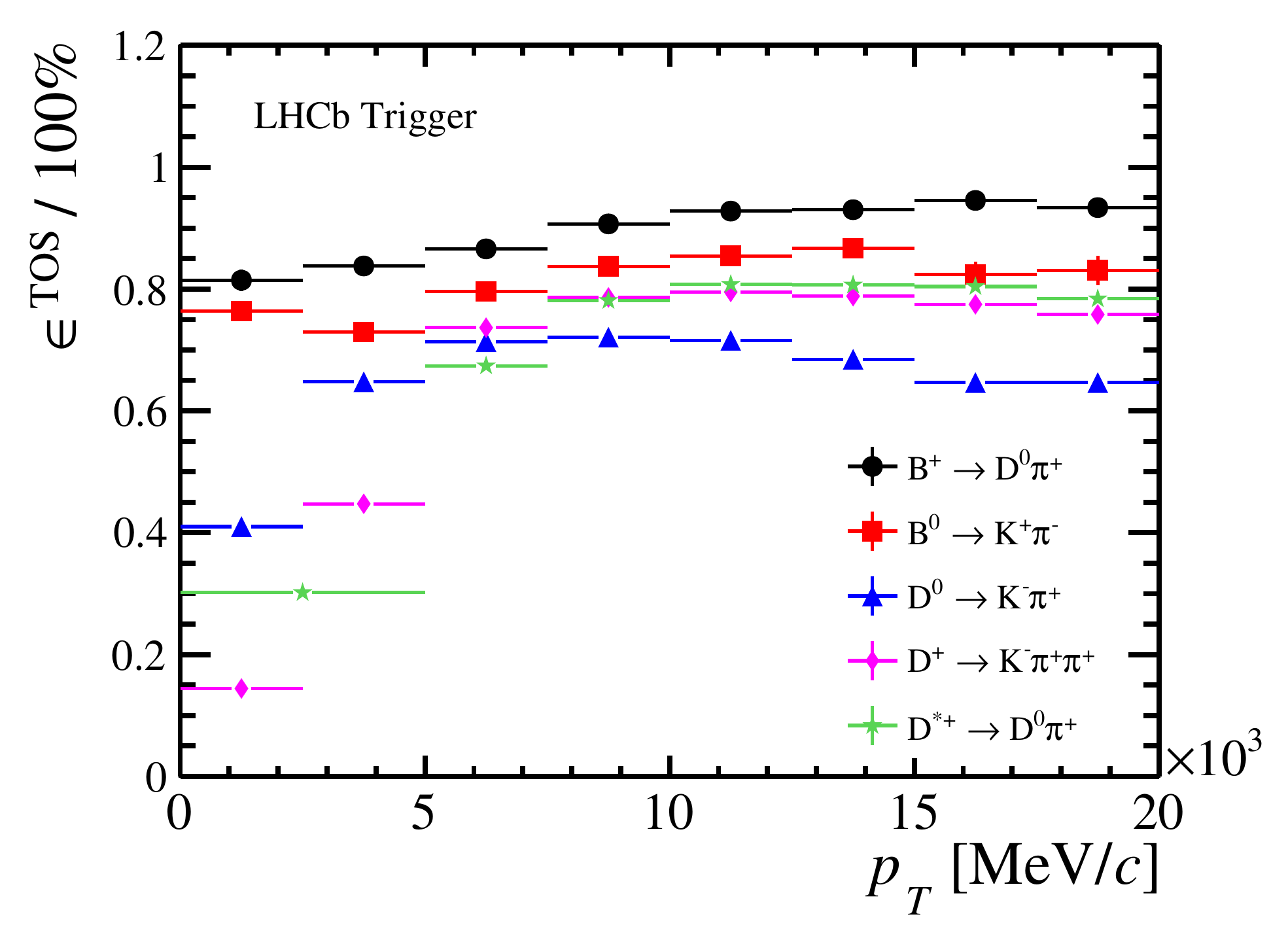}%
      \makebox[0cm][r]{\raisebox{0.22\textheight}[0cm]{\protect\subref{fig:hlt1:etos:pt}}\hspace{0.08\textwidth}}
    }%
    \subfloat{\label{fig:hlt1:etos:tau}%
      \includegraphics[type=pdf,ext=.pdf,read=.pdf,width=0.49\textwidth]{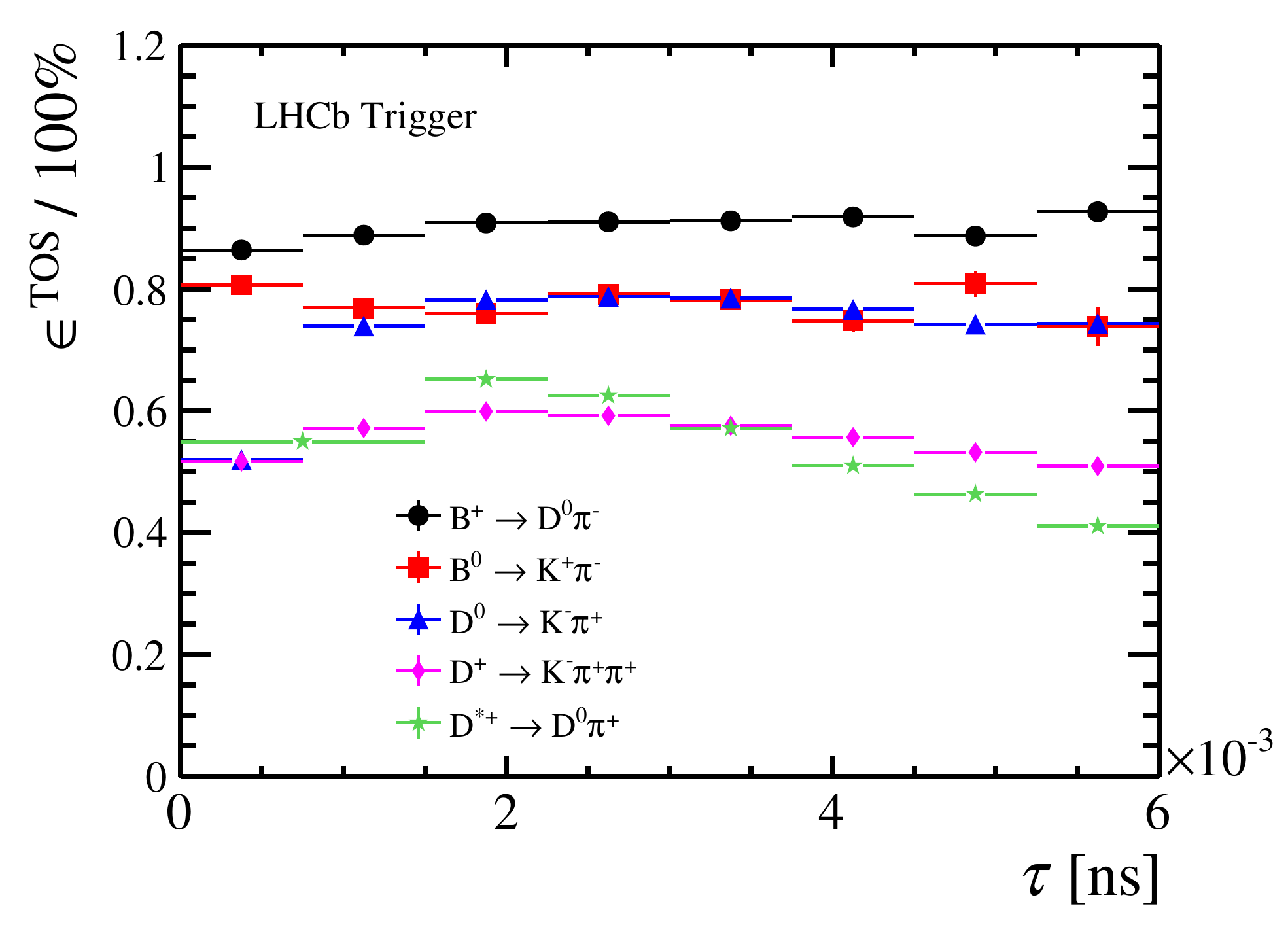}%
      \makebox[0cm][r]{\raisebox{0.22\textheight}[0cm]{\protect\subref{fig:hlt1:etos:tau}}\hspace{0.08\textwidth}}
    }

  \caption{\small The efficiency \hltonetrack for various reconstructed
        decay modes as functions of \protect\subref{fig:hlt1:etos:pt} \pT and
        \protect\subref{fig:hlt1:etos:tau} $\tau$ of the
        signal \PB or \PD candidate
        based on $\sqrt{s} = 8\TeV$ data collected in 2012.
        For the decay mode \DstarpTopipDzToKmpippippim,
        $\tau$ is the measured decay time of the \Dz candidate.
   \label{fig:hlt1:etos}}
  \end{figure}

  We evaluate the performance of \hltonetrack relative to the output of
  \lone with a set of offline candidate decays that are from events accepted
  by \lone and that are \TIS with respect at least one of the \hltone lines
  for physics analyses.
  \Figref{fig:hlt1:etos} shows \efftos of \hltonetrack as functions of \pT
  of the signal candidate and of measured decay time, $\tau$, of the
  signal \Dz or \Dp candidate for
  \DzToKmpip, \DpToKmpippip, and \DstarpTopipDzToKmpippippim decays.
  It also shows \efftos of \hltonetrack for two hadronic \PB decay modes,
  \decay{\Bp}{\pip\Dz(\Km\pip)} and \decay{\Bz}{\Kp\pim}.
  The mean efficiencies for the \lone-accepted \hltone-\TIS offline
  candidate decays appear in \Tabref{tab:hlt1:etos}.

  \begin{table}[hbt]
  \caption{\small Mean \efftos efficiencies of \hltonetrack relative to
        \lone-accepted events for selected charm hadron decays.
   \label{tab:hlt1:etos}}
    \centering
      \begin{tabular}{l|d{5}@{\,$\pm$\,}d{5}}
        Decay mode & \multicolumn{2}{c}{Mean \efftos} \\ \hline
        \DzToKmpip                      & 0.66853 & 0.00054 \\
        \DpToKmpippip                   & 0.58580 & 0.00014 \\
        \DstarpTopipDzToKmpippippim     & 0.60802 & 0.00038 \\
      \end{tabular}
  \end{table}

\section{\hlttwo performance}
\label{sec:hlt2}

  Like \hltone, \hlttwo is composed of several independent parallel lines,
  each of which is executed on each event accepted by at least one of the
  lines of \hltone.
  The decisions of each of the \lone channels and \hltone lines are available
  to \hlttwo and can enter the decision making of a line.
  Also like \hltone, most of the lines of \hlttwo begin with a common
  reconstruction of PVs and charged particle tracks.
  This reconstruction is more sophisticated, complete, and precise than that
  used by \hltone lines, but it also takes more computing power per event.
  Reference~\cite{LHCb-DP-2012-004} describes the \hlttwo reconstruction
  for data collection in 2011.
  Several improvements were made to the 2012 \hlttwo reconstruction, chief
  among them a reduction of the minimum \pT for reconstructed charged tracks
  from \mbox{$500\MeVc$} to \mbox{$300\MeVc$}.
  The \hlt deferral system provided the additional computational power
  necessary for this more complete track reconstruction.

  \subsection{Exclusive charm hadron lines}
  \label{sec:hlt2:charm}

    \hlttwo lines are generally tailored to the needs of groups of analyses.
    Because the precision and efficiency of \hlttwo's track reconstruction
    approach those of \LHCb's analysis software, \hlttwo lines can use the
    same methods and selection variables for fully reconstructing signal
    decays, with the exception of the charged hadron identification.
    The algorithms for the charged hadron identification require
    significant computational power and are executed only for a small number
    of \hlttwo lines on a relatively small number of events after extensive
    filtering.
    Among the lines for charm hadron physics, only the lines for \Lcp decays
    used the charged hadron identification.

\begin{figure}[tbhp]
    \centering
    \subfloat{\label{fig:hlt2:mass2:Kpi}%
      \includegraphics[type=pdf,ext=.pdf,read=.pdf,width=0.49\textwidth]{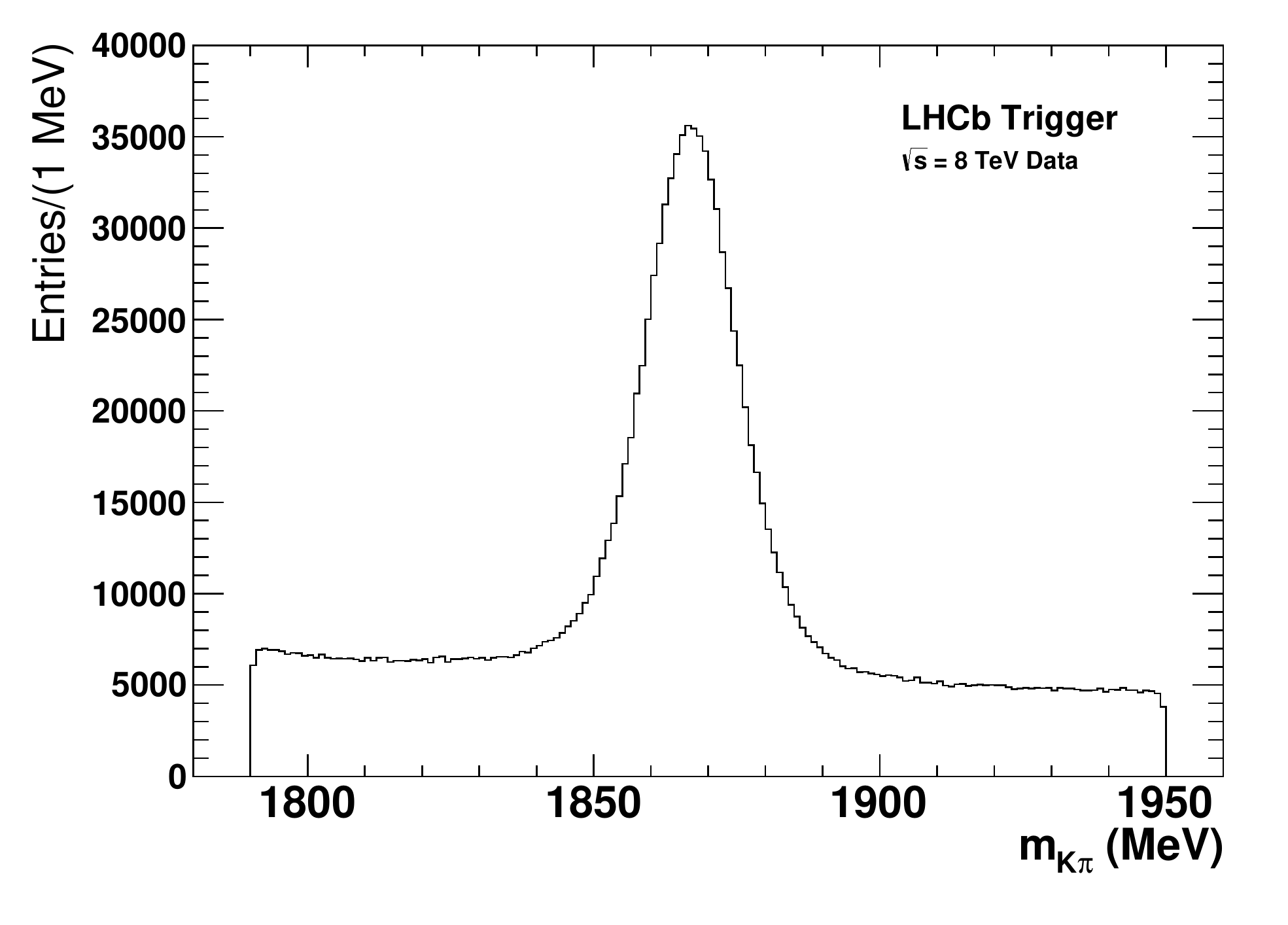}
      \makebox[0cm][r]{\raisebox{0.22\textheight}[0cm]{\protect\subref{fig:hlt2:mass2:Kpi}}\hspace{0.37\textwidth}}
    }%
    \subfloat{\label{fig:hlt2:mass3:3h}%
      \includegraphics[type=pdf,ext=.pdf,read=.pdf,width=0.49\textwidth]{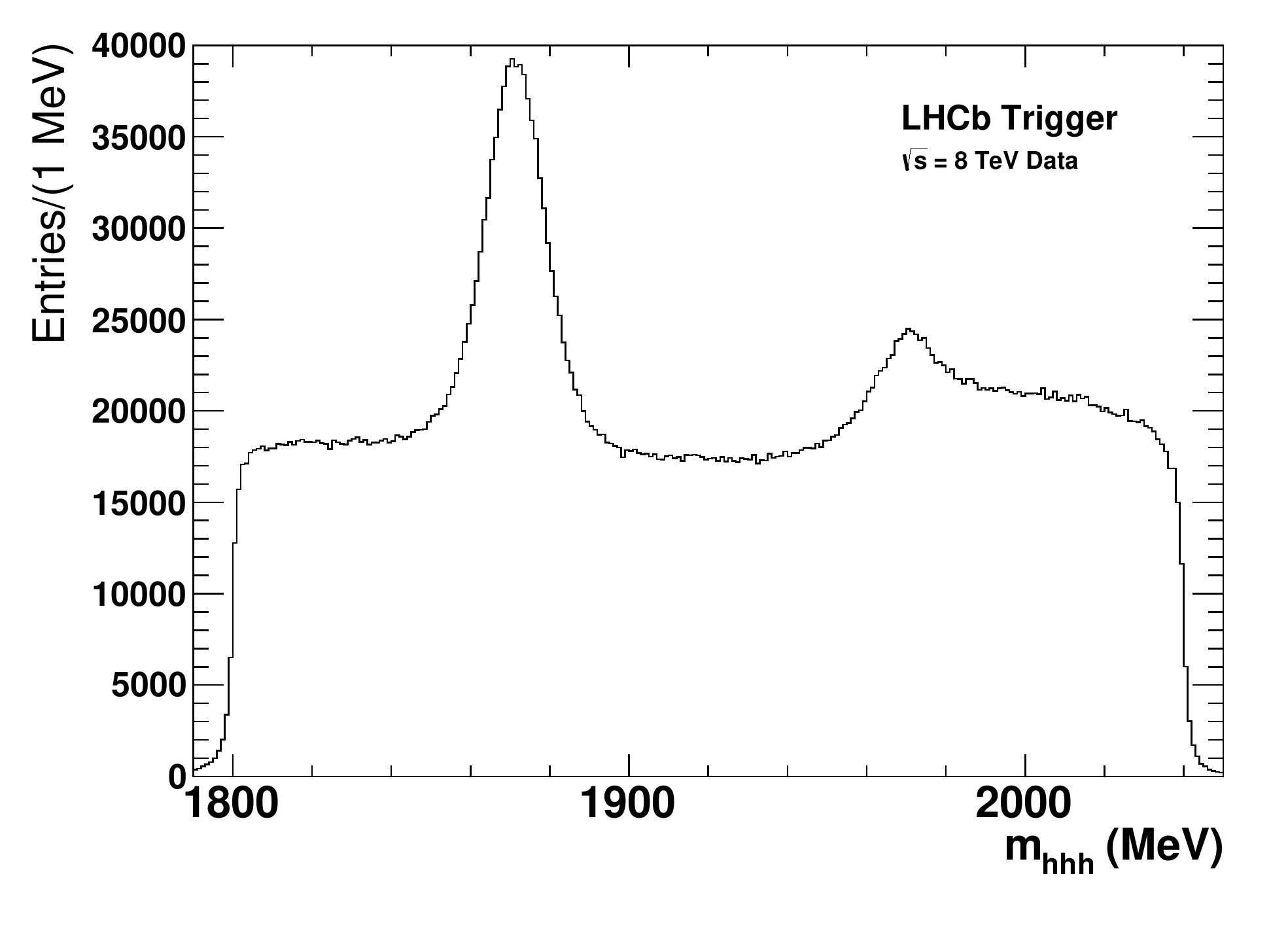}%
      \makebox[0cm][r]{\raisebox{0.22\textheight}[0cm]{\protect\subref{fig:hlt2:mass3:3h}}\hspace{0.37\textwidth}}
    }\\

    \subfloat{\label{fig:hlt2:mass3:4h}%
      \includegraphics[type=pdf,ext=.pdf,read=.pdf,width=0.49\textwidth]{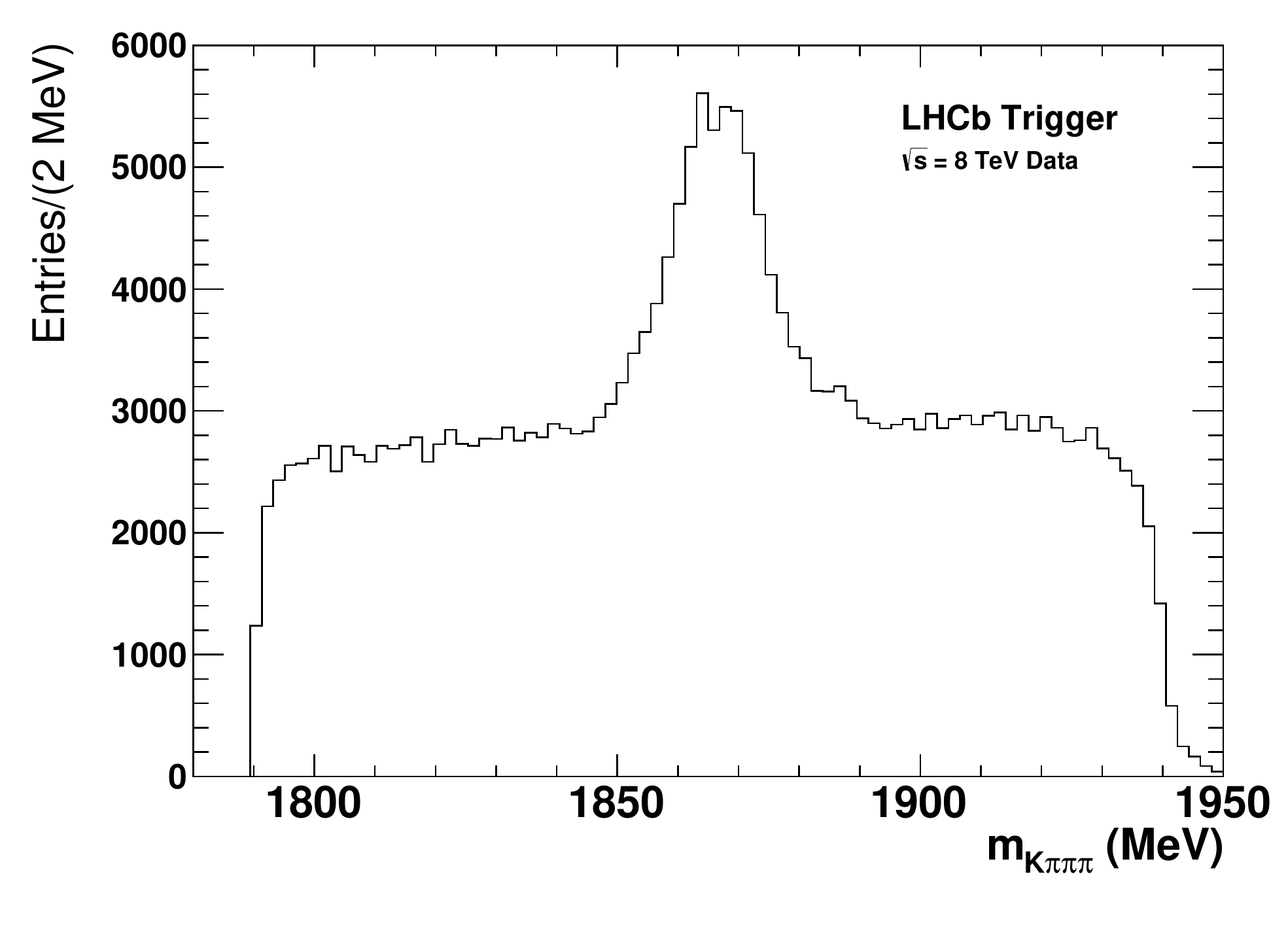}
      \makebox[0cm][r]{\raisebox{0.22\textheight}[0cm]{\protect\subref{fig:hlt2:mass3:4h}}\hspace{0.37\textwidth}}
    }

  \caption{\small Mass distributions of reconstructed \PD meson decay
        candidates in \hlttwo:
        \protect\subref{fig:hlt2:mass2:Kpi} \DzToKmpip candidates
        reconstructed in the line \protect\hlttwocharmtwobody,
        \protect\subref{fig:hlt2:mass3:3h} \decay{\Dpors}{\ensuremath{\Ph^{-}{\Ph'}^{+}{\Ph''}^{+}}}
        candidates, where
        \mbox{$\Ph, \Ph', \Ph'' \in \left\{ \kaon, \pion \right\}$},
        reconstructed in the line \protect\hlttwocharmthreebody, and
        \protect\subref{fig:hlt2:mass3:4h} \DzToKmpippippim
        candidates from the \DstarpTopipDz candidates reconstructed
        in the line \protect\hlttwocharmfourbody.
   \label{fig:hlt2:mass2}}
  \end{figure}

\begin{figure}[tbhp]
    \centering
    \subfloat{\label{fig:hlt2:etos:pt}%
      \includegraphics[type=pdf,ext=.pdf,read=.pdf,width=0.49\textwidth]{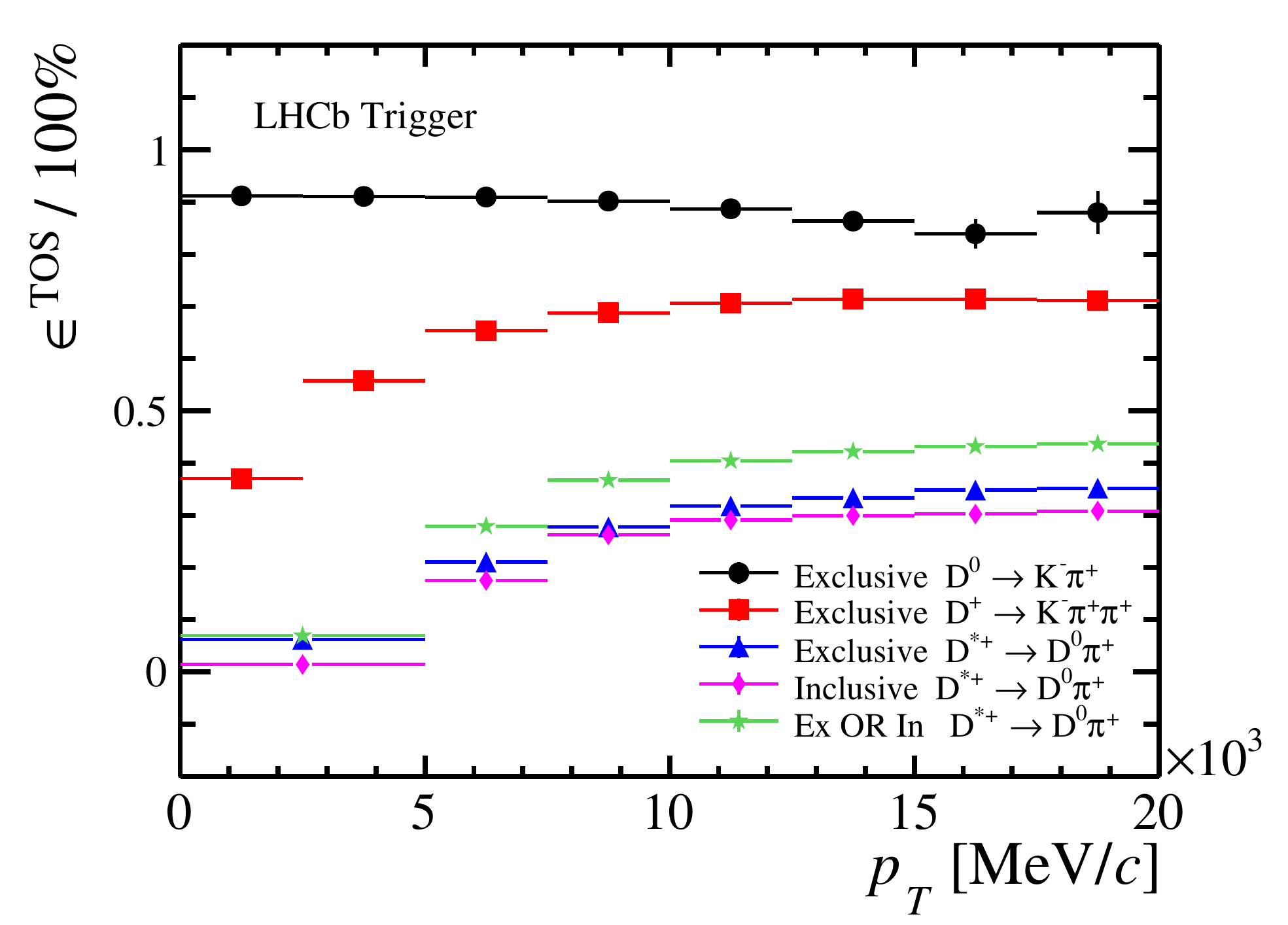}%
      \makebox[0cm][r]{\raisebox{0.22\textheight}[0cm]{\protect\subref{fig:hlt2:etos:pt}}\hspace{0.08\textwidth}}
    }%
    \subfloat{\label{fig:hlt2:etos:tau}%
      \includegraphics[type=pdf,ext=.pdf,read=.pdf,width=0.49\textwidth]{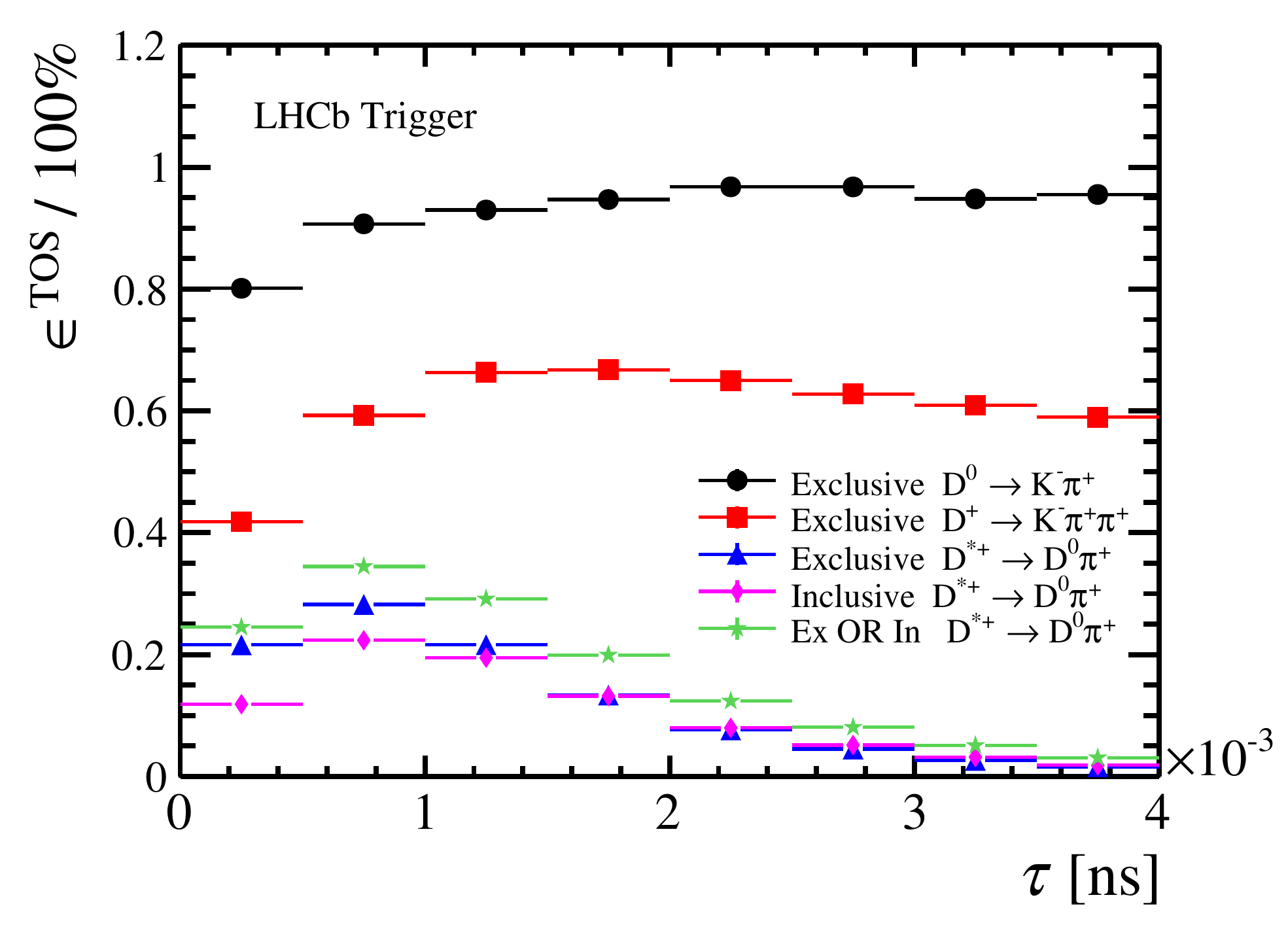}
      \makebox[0cm][r]{\raisebox{0.22\textheight}[0cm]{\protect\subref{fig:hlt2:etos:tau}}\hspace{0.08\textwidth}}
    }

  \caption{\small The efficiency of various \hlttwo lines for appropriate
        reconstructed decay modes as functions of
        \protect\subref{fig:hlt2:etos:pt} \pT and
        \protect\subref{fig:hlt2:etos:tau} $\tau$ of the signal \PD
        candidate based on $\sqrt{s} = 8\TeV$ data collected in 2012.
        For the decay mode \DstarpTopipDzToKmpippippim,
        $\tau$ is the measured decay time of the \Dz candidate.
   \label{fig:hlt2:etos}}
  \end{figure}

    The mass distributions of \Figref{fig:hlt2:mass2} demonstrate the purity
    with which charm hadron decays are reconstructed by their \hlttwo lines.
    We evaluate the performance of these lines relative to the output of
    \hltone with sets of offline candidate decays that are from events
    that are TOS with respect to one of the \hltone lines
    for physics  and that are \TIS with respect at least one of the \hlttwo
    lines for physics.
    \Figref{fig:hlt2:etos} shows \efftos of the \hlttwo lines as functions
    of \pT of the signal candidate and of measured decay time, $\tau$, of
    the signal \Dz or \Dp candidate.
    The mean efficiencies for the \lone-accepted \hltone-\TOS \hlttwo-TIS
    offline candidate decays appear in \Tabref{tab:hlt2:charm:etos}.

  \begin{table}[htp]
  \caption{\small Mean \efftos efficiencies of \hlttwo lines relative to
        \hltone-TOS events for selected charm hadron decays.
   \label{tab:hlt2:charm:etos}}
    \centering
      \begin{tabular}{ll|d{4}@{\,$\pm$\,}d{4}}
        Decay mode & \hlttwo line & \multicolumn{2}{c}{Mean \efftos} \\ \hline
        \DzToKmpip
        & \hlttwocharmtwobody
        & 0.9069 & 0.0015 \\

        \DpToKmpippip
        & \hlttwocharmthreebody
        & 0.6588 & 0.0005 \\

        \DstarpTopipDzToKmpippippim
        & \hlttwocharmfourbody
        & 0.1989 & 0.0004 \\

        & \hlttwocharmincldstar
        & 0.1712 & 0.0005 \\

        & \hlttwocharmfourbody 
        & \multicolumn{2}{c}{} \\
        
        & \hspace{0.5em}or \hlttwocharmincldstar
        & 0.2556 & 0.0005 \\
      \end{tabular}
  \end{table}

\clearpage

  \subsection{Inclusive \Dstarp line}
  \label{sec:hlt2:dstar}

  \begin{wrapfigure}{R}{0.5\textwidth}
    \centering
     \includegraphics[type=pdf,ext=.pdf,read=.pdf,width=0.49\textwidth]{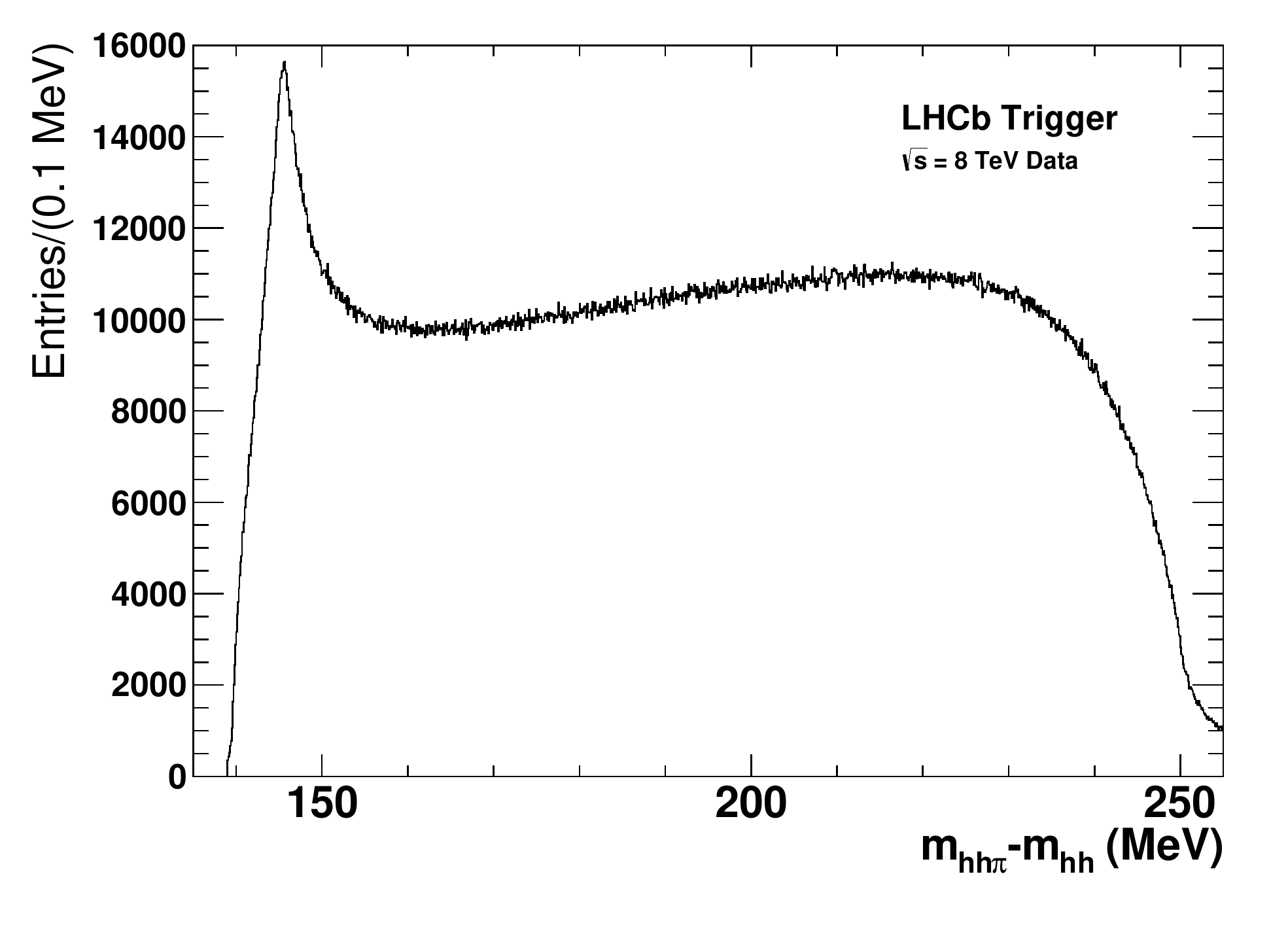}
  \caption{\small Mass difference distribution of reconstructed candidates
        for the \hlttwo inclusive \Dstarp line \hlttwocharmincldstar.
   \label{fig:hlt2:massincl}}
  \end{wrapfigure}

  Although highly successful, \hlttwo lines for exclusive reconstruction
  of decay modes are necessarily limited.
  Inclusive selections that do not depend on a complete reconstruction of
  signal decays can allow for efficient selection of a broader range of decay
  modes, including modes for which a full reconstruction is impossible.
  The inclusive \Dstarp \hlttwo line is a first example of inclusive
  triggering for charm hadrons.

  The inclusive \Dstarp line,\\
  \hlttwocharmincldstar, selects decays of
  \DstarpTopipDz, where \Dz decays into at least two charged final state
  particles.
  Partial \Dz decay candidates are reconstructed as two-track vertexes
  that are significantly displaced from all PVs.
  These two-track vertexes are combined with \pip candidates to form \Dstarp
  candidates, and additional basic kinematic and reconstruction quality
  criteria are applied to the system.
  For true \Dstarp decays, the mass difference between the reconstructed
  \Dstarp and \Dz candidates peaks strongly at the true value, even when the
  \Dz decays are not fully reconstructed.
  Thus \Dstarp decays can be successfully identified for a wide array of
  \Dz decay modes.
  The method has also been applied to
  \decay{\ensuremath{\PSigma_{\cquark}^{0(++)}}}{\ensuremath{\Lcp \pion^{-(+)}}}
  decay modes with partially reconstructed \Lcp decays
  in additional \hlttwo lines.
  \Figref{fig:hlt2:massincl} shows the prominent signal component in the
  \Dstarp-\Dz candidate mass differences for the \Dstarp candidates selected by
  the inclusive \Dstarp \hlttwo line.

  \Figref{fig:hlt2:etos} includes a comparison of the performance of the
  inclusive \Dstarp line with that of the exclusive line for
  \DstarpTopipDzToKmpippippim decays.
  The inclusive line has a comparable efficiency and, furthermore,
  selects a complementary set of decays as can be seen in the efficiencies
  of \Tabref{tab:hlt2:charm:etos}.
  Approximately 33\% of the signal decays selected by the inclusive line were
  not selected by the exclusive line.
  Most of these are decays for which one of the final state particles
  has $\pT < 300\MeVc$, the lower limit for the track reconstruction in the
  exclusive lines.

\section{Future developments}
\label{sec:future}

  \subsection{Post-LS1 \LHCb triggering}
  \label{sec:future:ls1}

    In 2015 \LHCb will resume data collection after \lhc's LS1 at the greater
    $\proton\proton$ collision energy of $\sqrt{s} = 13\TeV$.
    The \lone hardware trigger will be tuned to satisfy its $1\MHz$ output
    limit under the new conditions, but its operation will remain unchanged.
    The \hlt software trigger will be substantially reorganized in order to
    improve the quality of the event reconstruction in \hlttwo.

    The internal structures of \hltone and \hlttwo will remain largely
    unchanged, with the possible addition of lines to expand \LHCb's
    physics program.
    However, an additional calibration step will be inserted between
    \hltone and \hlttwo.
    In 2010-2012, the calibration and the fine alignment of detector elements
    that was used by the \hlttwo reconstruction were measured in an earlier
    data-taking period.
    Since the calibration and alignment for analysis is always up-to-date,
    there may be small differences between the measured parameters of identical
    candidates as reconstructed in \hlttwo and as reconstructed for analysis.
    This can be a source of irreducible systematic uncertainty.
    By performing the calibration and alignment step before the execution of
    \hlttwo, this source of uncertainty is reduced or eliminated.

    \hltone will run immediately on all \lone-accepted events.
    The events accepted by \hltone will be cached on the storage of the EFF
    by a system similar to the \hlt deferral until an update of
    the a detector alignment and calibration is complete.
    Then \hlttwo will process the cached \hltone-accepted events and
    render the final trigger decisions.

  \subsection{Triggering in an upgraded \LHCb detector}
  \label{sec:future:upgrade}

    Following the conclusion of \lhc Run II, the \LHCb experiment will be
    upgraded for a higher rate of data
    collection~\cite{CERN-LHCC-2011-001,Bediaga:1443882}.
    The upgraded experiment will feature a substantially improved trigger.
    Inefficiency in the \lone trigger is one of the main limitations of the
    current system for \bquark and \cquark-hadron decays to hadronic final
    states.
    This inefficiency is necessitated by $1\MHz$ maximum readout rate for
    the detector electronics.
    The upgraded \LHCb detector will be capable of a full detector readout at
    $40\MHz$, largely obviating the need for \lone.
    \lone will be upgraded to a Low Level Trigger that will function as
    a pass-through during normal operation.
    All trigger decisions will be made by the more flexible and efficient
    \hlt, which will evolve to process the higher input rate.
    The rate at which events are accepted by the trigger for permanent storage
    will increase from the current $5\kHz$ to an estimated $20\kHz$.
    The combination of a more efficient software trigger and the increased
    rate of data collection is estimated to increase the annual yield of
    many charm decay modes by an order of magnitude.

\section{Summary}
\label{sec:summary}

  The current performance of the \LHCb charm triggering, as documented in
  this article, is the product of steady iterative improvement made with the
  goal of expanding the scope and impact of \LHCb's physics program.
  Development of the trigger system continues, with further important
  enhancements anticipated for \lhc Run II and for the subsequent upgrade
  of the \LHCb experiment.
  The \LHCb trigger will continue to deliver world-class charm data sets
  for many years.


\begin{mcitethebibliography}{10}
\mciteSetBstSublistMode{n}
\mciteSetBstMaxWidthForm{subitem}{\alph{mcitesubitemcount})}
\mciteSetBstSublistLabelBeginEnd{\mcitemaxwidthsubitemform\space}
{\relax}{\relax}

\bibitem{LHCb-PAPER-2013-053}
LHCb collaboration, R.~Aaij {\em et~al.},
  \ifthenelse{\boolean{articletitles}}{{\it {Measurement of $\Dz$--$\Dzb$
  mixing parameters and search for \CP violation using $\Dz \to \Kp\pim$
  decays}}, }{}\href{http://arxiv.org/abs/1309.6534}{{\tt arXiv:1309.6534}},
  {submitted to Phys. Rev. Lett.}\relax
\mciteBstWouldAddEndPunctfalse
\mciteSetBstMidEndSepPunct{\mcitedefaultmidpunct}
{}{\mcitedefaultseppunct}\relax
\EndOfBibitem
\bibitem{LHCb-PAPER-2013-037}
LHCb collaboration, R.~Aaij {\em et~al.},
  \ifthenelse{\boolean{articletitles}}{{\it {Measurement of form factor
  independent observables in the decay $B^0\to K^{*0}\mu^+\mu^-$}},
  }{}\href{http://arxiv.org/abs/1308.1707}{{\tt arXiv:1308.1707}}, {to appear
  in Phys. Rev. Lett.}\relax
\mciteBstWouldAddEndPunctfalse
\mciteSetBstMidEndSepPunct{\mcitedefaultmidpunct}
{}{\mcitedefaultseppunct}\relax
\EndOfBibitem
\bibitem{LHCb-PAPER-2010-002}
LHCb collaboration, R.~Aaij {\em et~al.},
  \ifthenelse{\boolean{articletitles}}{{\it {Measurement of $\sigma(pp \to
  \bbbar X)$ at $\sqrt{s}=7\tev$ in the forward region}},
  }{}\href{http://dx.doi.org/10.1016/j.physletb.2010.10.010}{Phys.\ Lett.\
  {\bf B694} (2010) 209}, \href{http://arxiv.org/abs/1009.2731}{{\tt
  arXiv:1009.2731}}\relax
\mciteBstWouldAddEndPuncttrue
\mciteSetBstMidEndSepPunct{\mcitedefaultmidpunct}
{\mcitedefaultendpunct}{\mcitedefaultseppunct}\relax
\EndOfBibitem
\bibitem{LHCb-PAPER-2012-041}
LHCb collaboration, R.~Aaij {\em et~al.},
  \ifthenelse{\boolean{articletitles}}{{\it {Prompt charm production in $pp$
  collisions at $\sqrt s= 7\tev$}},
  }{}\href{http://dx.doi.org/10.1016/j.nuclphysb.2013.02.010}{Nucl.\ Phys.\
  {\bf B871} (2013) 1}, \href{http://arxiv.org/abs/1302.2864}{{\tt
  arXiv:1302.2864}}\relax
\mciteBstWouldAddEndPuncttrue
\mciteSetBstMidEndSepPunct{\mcitedefaultmidpunct}
{\mcitedefaultendpunct}{\mcitedefaultseppunct}\relax
\EndOfBibitem
\bibitem{Alves:2008zz}
LHCb collaboration, A.~A. Alves~Jr. {\em et~al.},
  \ifthenelse{\boolean{articletitles}}{{\it {The \lhcb detector at the LHC}},
  }{}\href{http://dx.doi.org/10.1088/1748-0221/3/08/S08005}{JINST {\bf 3}
  (2008) S08005}\relax
\mciteBstWouldAddEndPuncttrue
\mciteSetBstMidEndSepPunct{\mcitedefaultmidpunct}
{\mcitedefaultendpunct}{\mcitedefaultseppunct}\relax
\EndOfBibitem
\bibitem{LHCb-DP-2012-003}
M.~Adinolfi {\em et~al.}, \ifthenelse{\boolean{articletitles}}{{\it
  {Performance of the \lhcb RICH detector at the LHC}},
  }{}\href{http://dx.doi.org/10.1140/epjc/s10052-013-2431-9}{Eur.\ Phys.\ J.\
  {\bf C73} (2013) 2431}, \href{http://arxiv.org/abs/1211.6759}{{\tt
  arXiv:1211.6759}}\relax
\mciteBstWouldAddEndPuncttrue
\mciteSetBstMidEndSepPunct{\mcitedefaultmidpunct}
{\mcitedefaultendpunct}{\mcitedefaultseppunct}\relax
\EndOfBibitem
\bibitem{LHCb-DP-2012-002}
A.~A. Alves~Jr. {\em et~al.}, \ifthenelse{\boolean{articletitles}}{{\it
  {Performance of the LHCb muon system}},
  }{}\href{http://dx.doi.org/10.1088/1748-0221/8/02/P02022}{JINST {\bf 8}
  (2013) P02022}, \href{http://arxiv.org/abs/1211.1346}{{\tt
  arXiv:1211.1346}}\relax
\mciteBstWouldAddEndPuncttrue
\mciteSetBstMidEndSepPunct{\mcitedefaultmidpunct}
{\mcitedefaultendpunct}{\mcitedefaultseppunct}\relax
\EndOfBibitem
\bibitem{LHCbTDR:2003tg}
LHCb collaboration, R.~Antunes-Nobrega {\em et~al.},
  \ifthenelse{\boolean{articletitles}}{{\it {LHCb} trigger system : Technical
  design report}, }{} \href{http://cds.cern.ch/record/630828}{CERN-LHCC-2003-031} (2003),
\newblock LHCb TDR 10\relax
\mciteBstWouldAddEndPuncttrue
\mciteSetBstMidEndSepPunct{\mcitedefaultmidpunct}
{\mcitedefaultendpunct}{\mcitedefaultseppunct}\relax
\EndOfBibitem
\bibitem{LHCb-DP-2012-004}
R.~Aaij {\em et~al.}, \ifthenelse{\boolean{articletitles}}{{\it {The \lhcb
  trigger and its performance in 2011}},
  }{}\href{http://dx.doi.org/10.1088/1748-0221/8/04/P04022}{JINST {\bf 8}
  (2013) P04022}, \href{http://arxiv.org/abs/1211.3055}{{\tt
  arXiv:1211.3055}}\relax
\mciteBstWouldAddEndPuncttrue
\mciteSetBstMidEndSepPunct{\mcitedefaultmidpunct}
{\mcitedefaultendpunct}{\mcitedefaultseppunct}\relax
\EndOfBibitem
\bibitem{Adeva:2009ny}
LHCb collaboration, B.~Adeva {\em et~al.},
  \ifthenelse{\boolean{articletitles}}{{\it {Roadmap for selected key
  measurements of LHCb}}, }{}\href{http://arxiv.org/abs/0912.4179}{{\tt
  arXiv:0912.4179}}\relax
\mciteBstWouldAddEndPuncttrue
\mciteSetBstMidEndSepPunct{\mcitedefaultmidpunct}
{\mcitedefaultendpunct}{\mcitedefaultseppunct}\relax
\EndOfBibitem
\bibitem{LHCb-PAPER-2012-031}
LHCb collaboration, {R. Aaij, \emph{et al.}, and A. Bharucha} {\em et~al.},
  \ifthenelse{\boolean{articletitles}}{{\it {Implications of LHCb measurements
  and future prospects}},
  }{}\href{http://dx.doi.org/10.1140/epjc/s10052-013-2373-2}{Eur.\ Phys.\ J.\
  {\bf C73} (2013) 2373}, \href{http://arxiv.org/abs/1208.3355}{{\tt
  arXiv:1208.3355}}\relax
\mciteBstWouldAddEndPuncttrue
\mciteSetBstMidEndSepPunct{\mcitedefaultmidpunct}
{\mcitedefaultendpunct}{\mcitedefaultseppunct}\relax
\EndOfBibitem
\bibitem{Evans:2008zzb}
L.~Evans and P.~Bryant, \ifthenelse{\boolean{articletitles}}{{\it {LHC
  Machine}}, }{}\href{http://dx.doi.org/10.1088/1748-0221/3/08/S08001}{JINST
  {\bf 3} (2008) S08001}\relax
\mciteBstWouldAddEndPuncttrue
\mciteSetBstMidEndSepPunct{\mcitedefaultmidpunct}
{\mcitedefaultendpunct}{\mcitedefaultseppunct}\relax
\EndOfBibitem
\bibitem{CERN-LHCC-2011-001}
LHCb collaboration, R.~Aaij {\em et~al.},
  \ifthenelse{\boolean{articletitles}}{{\it {Letter of Intent for the LHCb
  Upgrade}}, }{} \href{http://cds.cern.ch/record/1333091}{CERN-LHCC-2011-001}, LHCC-I-018 (2011)\relax
\mciteBstWouldAddEndPuncttrue
\mciteSetBstMidEndSepPunct{\mcitedefaultmidpunct}
{\mcitedefaultendpunct}{\mcitedefaultseppunct}\relax
\EndOfBibitem
\bibitem{Bediaga:1443882}
LHCb collaboration, I.~Bediaga {\em et~al.},
  \ifthenelse{\boolean{articletitles}}{{\it {Framework TDR for the LHCb
  Upgrade: Technical Design Report}}, }{} \href{http://cds.cern.ch/record/1443882}{CERN-LHCC-2012-007},
  LHCb-TDR-12 (2012)\relax
\mciteBstWouldAddEndPuncttrue
\mciteSetBstMidEndSepPunct{\mcitedefaultmidpunct}
{\mcitedefaultendpunct}{\mcitedefaultseppunct}\relax
\EndOfBibitem
\end{mcitethebibliography}


\ifx\mcitethebibliography\mciteundefinedmacro
\PackageError{LHCb.bst}{mciteplus.sty has not been loaded}
{This bibstyle requires the use of the mciteplus package.}\fi
\providecommand{\href}[2]{#2}

\end{document}